# Gradient descent-based programming of analog in-memory computing cores

J. Büchel[1], A. Vasilopoulos[1], B. Kersting[1], F. Odermatt[1], K. Brew[2], I. Ok[2], S. Choi[2], I. Saraf[2]
V. Chan[2], T. Philip[2], N. Saulnier[2], V. Narayanan[3], M. Le Gallo[1], A. Sebastian[1]
[1]IBM Research Europe, Rüschlikon, Switzerland, email: jub@zurich.ibm.com, ase@zurich.ibm.com
[2]IBM Research – Albany, NY, USA   [3]IBM Research – Yorktown Heights, NY, USA

*Abstract*— The precise programming of crossbar arrays of unit-cells is crucial for obtaining high matrix-vector-multiplication (MVM) accuracy in analog in-memory computing (AIMC) cores. We propose a radically different approach based on directly minimizing the MVM error using gradient descent with synthetic random input data. Our method significantly reduces the MVM error compared with conventional unit-cell by unit-cell iterative programming. It also eliminates the need for high-resolution analog-to-digital converters (ADCs) to read the small unit-cell conductance during programming. Our method improves the experimental inference accuracy of ResNet-9 implemented on two phase-change memory (PCM)-based AIMC cores by 1.26%.

## I. Introduction

AIMC based on crossbar arrays of unit-cells comprising memristive devices promises significant gains in latency and power efficiency for a wide range of applications [1,2]. By applying voltages to the crossbar rows and reading out the currents that accumulate along the columns, one can perform MVMs in constant time complexity. However, the MVM is imprecise, affecting the accuracy of all downstream tasks, such as neural network inference. The MVM error is mostly attributed to the conductance variations of the memristive devices. The programming error (inaccuracy associated with mapping a synaptic weight to a unit-cell) is a key component along with conductance drift and 1/f noise [3,4]. Besides the conductance variations which can be treated as the linear component of the MVM error, there are also nonlinearities attributed to the ADCs, IR-drop, etc.

The current state-of-the-art algorithm [5] for programming crossbar arrays relies on iteratively reading and correcting the individual unit-cell conductance until the read-out value is within a predefined error margin of the target weight (Fig. 1a). Once a device is considered converged, it is typically disregarded for the rest of the programming procedure. Hence, the device conductance may drift away from its target value while the other devices are programmed, leading to additional uncorrected errors. Moreover, accurately reading the conductance of an individual device requires a highly precise read circuitry as well as long integration times, since currents <1$\mu$A are sensed. The ADCs must support a wide dynamic range because they need to read the current that is generated by a single device with high precision, as well as the current generated by a whole column of high conductance devices. Therefore, accurate reading of such small currents is typically not possible with the integrated read circuitry of an AIMC core, and the programming accuracy will be ultimately limited by the error incurred during the reading process [3]. This problem is further accentuated when using memristive devices with smaller conductance values, which are beneficial for reduced power consumption and IR-drop. In the context of neural network inference, there have been proposals of partially re-tuning the weights using chip-in-the-loop training after the network has been programmed on the chip to correct such programming errors [6]. However, this requires implementing the gradient calculations that arise from backpropagation, introduces network-specific overhead when mapping pre-trained models to the chip and additional memory costs due to the use of the training data.

To alleviate those issues, we propose a novel programming scheme that relies on directly optimizing the MVM accuracy per core instead of the individual device conductance values. By doing so, we can significantly reduce the MVM error with respect to iterative programming without introducing network-specific overheads in the programming scheme.

## II. Gradient descent-based programming

In **G**radient **D**escent-based **P**rogramming (GDP), after initializing the unit-cell conductances, we perform batched MVMs on-chip with randomly generated input vectors. The resulting MVM error is quantified as a loss function. We then calculate gradients of this loss function and use them to set the amplitudes of the programming pulses sent to the chip to update the devices (see Fig. 1b and Fig. 1c). This procedure is iterated until a satisfactory MVM error is reached. Hence, the only operation that needs to be supported by the chip is the MVM and accurate single-device reads are not needed. Furthermore, all unit-cells can be updated at every iteration, eliminating the possibility that some weights might start to diverge while the other devices are still being programmed.

## III. MVM characterization methodology

To quantify the accuracy achieved with GDP, we developed a methodology to characterize MVMs from AIMC cores. Given a target weight matrix $G$ and input matrix $x$, the exact MVM is denoted $y = Gx$. The result when the same computation is done on the AIMC core is denoted $\tilde{y}$. We can solve $\hat{G} = \text{argmin}_{\hat{G}} \, ||\tilde{y} - \hat{G}x\,||_2$ to obtain an estimation of the weights $\hat{G}$ that have been programmed onto the array. We can then calculate the exact result of the MVM performed with the estimated weights: $\hat{y} = \hat{G}x$. Using these values, we define various error metrics (see Fig. 2 for the definitions). The weight error $\epsilon_{\text{weight}}$ indicates how close the programmed weights are to the target weights. The total MVM error $\epsilon_{\text{total}}$ quantifies the MVM error that we minimize, and the nonlinear error $\epsilon_{\text{nonlinear}}$ captures the residual error mostly arising from the nonlinearities of the peripheral circuitry, IR-drop etc.

## IV. Experimental Results

All experiments were carried out on two 14-nm PCM-based AIMC cores [7], where each core is equipped with a $256 \times 256$ crossbar array (Fig. 3). Each unit-cell in the crossbar comprises four PCM devices (two for each polarity).

First, we investigate how devices in the crossbar need to be initialized to roughly represent the target conductance matrix before running GDP. This can be achieved by either running the conventional iterative algorithm for some steps or by using single shot programming with pulse amplitudes being a function of the target conductances. As Fig. 4 shows, both initialization schemes work equally well, and subsequently running GDP significantly improves the MVM accuracy over the conventional iterative programming approach (see Fig. 5). This improvement is partly because GDP also slightly corrects for nonlinearities introduced by the hardware. Fig. 6 illustrates that GDP, in contrast to the iterative method, converges to readout weights that are slightly different from the target weights, which lead to a lower MVM error.

Fig. 5 shows results obtained with one device per polarity in the unit-cell. However, using two devices per polarity can be beneficial because it increases the signal-to-noise ratio (SNR) [8]. When using two devices, we program the first device to either SET or RESET state and run GDP on the other (Fig. 7). As Fig. 8 illustrates, GDP consistently lowers the MVM error compared to the iterative approach when both one and two devices are used. Moreover, running two-device GDP for 500 iterations allows to reach the noise floor given by the nonlinear error. The programming error eventually vanishes over iterations and the total error is solely dominated by the nonlinearities of the system.

Having established that GDP improves on the iterative approach right after programming, we investigated whether this improvement was also retained over time while the PCM devices drift. Fig. 9 shows the total and nonlinear errors over a period of 24 hours for the different programming schemes. A side-by-side comparison of the weight error between single-device iterative programming and GDP is shown in Fig. 10. GDP consistently retains the weight error reduction on the iterative approach over time.

We also performed experiments on a second type of PCM device with lower maximum conductance (see Fig. 11). As a result of the low conductance and limited resolution of the ADCs, the single device reads are imprecise, and the iterative algorithm yields high MVM errors. However, GDP allows to reach comparable MVM errors to those achieved on higher conductive PCM because it is not reliant on single device reads.

For GDP to be a viable substitution for the iterative programming algorithm, we must make sure that the inputs needed for GDP are not weight/application-dependent but can be generated with a random number generator, since using specific data for programming would incur additional memory requirements. We therefore tested whether GDP still outperforms the iterative approach on data that is different from the data that was used during programming. Fig. 12 shows that this is the case, even when the sparsity or the whole distribution of the input data changes. In addition to being agnostic to the input data, GDP should also be robust to the choice of hyper-parameters. The only hyperparameter that GDP introduces is the learning rate that is used to scale the weight updates to guarantee smooth convergence. Fig. 13 shows that good performance is achieved if the learning rate is large enough and its precise value does not matter.

GDP also exhibits reasonable computational and memory complexity. Unit cells can be written and read linearly in the number of rows $r$ of a crossbar, while the MVM can be performed in constant time complexity. Because GDP avoids reading devices, we effectively have only one linear dependency on $r$ (for programming) and are not limited by the read time. However, GDP relies on digitally computing the gradient, which is $\mathcal{O}(Brc)$, where $B$ is the batch size and $c$ is the number of columns. The iterative algorithm and GDP both have a memory complexity of $\mathcal{O}(rc)$. Fig. 14 shows the performance of GDP for varying batch sizes of the input data. GDP outperforms the iterative approach at a batch size of 64 and best results are achieved with sizes of 256 or higher (larger than the number of columns).

We finally investigated whether our method could improve the CIFAR-10 inference accuracy with ResNet-9 implemented on two PCM cores. The layers were programmed on both cores, all MVMs were performed on-chip and other computations in software (Fig. 15). Fig. 16 shows that GDP produces lower per-layer MVM errors and improves the inference accuracy from 84.75% to 86.01%.

## V. Conclusion

We demonstrate a novel gradient-descent based scheme for programming unit cells in an AIMC core performing MVMs. Compared with the conventional iterative programming algorithm, we achieve superior performance in terms of MVM accuracy and retain it over time. Our method eliminates the need for ADCs that are designed to read out single devices and facilitates the use of highly resistive memristive devices for AIMC. Our algorithm does not require extra memory for storing inputs and is robust to the choice of the learning rate. Finally, we experimentally demonstrate that the neural network inference accuracy is directly improved by using this approach.

This work was supported by the IBM Research AI Hardware Center. This work has also received funding from the European Union's Horizon Europe research and innovation programme under Grant Agreement No 101046878, and was supported by the Swiss State Secretariat for Education, Research and Innovation (SERI) under contract number 22.00029.

References

[1] Lanza et al., Science 376, 1066 (2022)
[2] Yu et al., IEEE Circuits Syst. Mag. 21, pp. 31-56 (2021)
[3] Nandakumar et al., IEDM Tech. Dig., pp. 29.4.1-4 (2020)
[4] Mackin et al., Nature Communications, v. 13, 3765 (2022)
[5] Papandreou et al., Proc. ISCAS, pp. 329-332 (2011)
[6] Yao et al., Nature 577, pp. 641-646 (2020)
[7] Khaddam-Aljameh et al., JSSC 57 (4), pp. 1027-1038 (2022)
[8] Le Gallo et al., Neuromorphic Comput. Eng. 2, 014009 (2022)

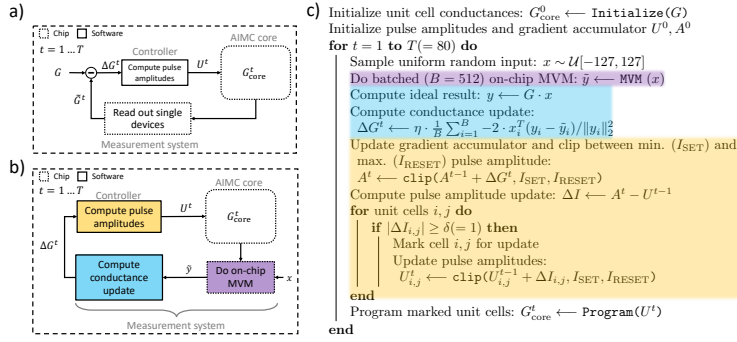

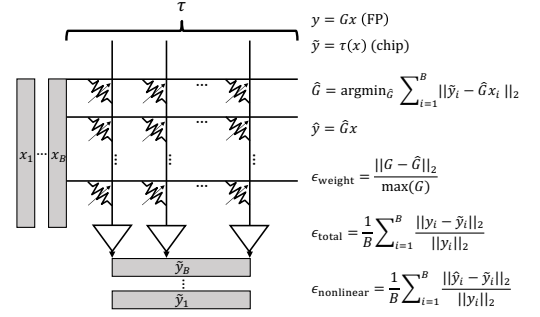

Fig. 1. (a) In standard iterative programming, conductance values of the cros out ($\tilde{G}$) and pulses ($U$) proportional to the difference between the reado conductances ($G$) are used to update the crossbar conductances ($G_{core}$). descent-based programming (GDP) uses gradients of the MVM error programming pulses that iteratively bring the crossbar conductances cl targets. (c) Pseudocode of GDP.

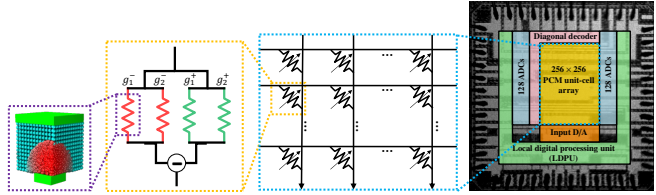

Fig. 3. For the experiments, we use two fully-integrated PCM chips [7] with 256x256 unit cells per chip. Each unit cell comprises four PCM devices (two for each polarity).

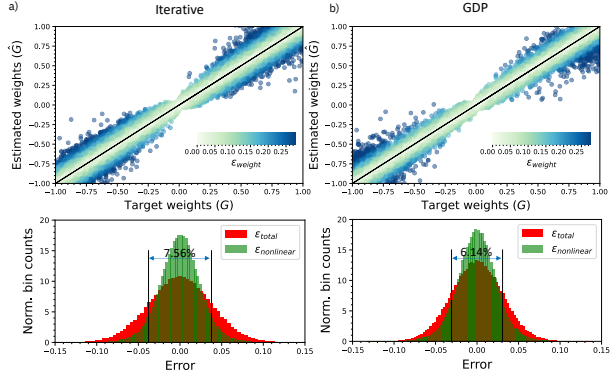

Fig. 5. Comparison between the iterative programming algorithm (a) and GDP (b). GDP reduces the per unit-cell weight error and total MVM error.

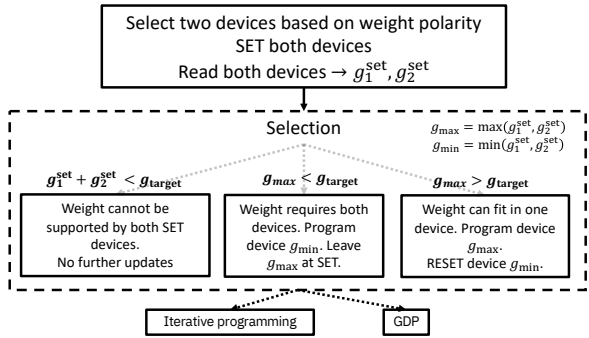

Fig. 7. Programming algorithm when using two devices per polarity. If the weight fits on just one device, we choose the one with highest SET conductance for programming and RESET the other. If both devices cannot accommodate the weight, we do not program the weight and leave both devices in SET state. Finally, if both devices are needed to fit the weight, we choose the device with lowest SET conductance for programming. The conductance of a single device is denoted $g$.

quantifies the per-unit-cell programming error.

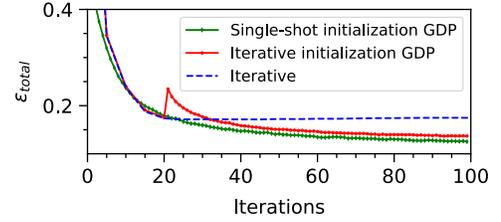

Fig. 4. Weights for GDP can be initialized by either running the iterative algorithm for some (here 20) iterations (red) or by applying a single-shot RESET pulse to the devices (green), where the pulse amplitudes are a function of the target values. The jump in the error of GDP after the iterative initialization (20 iterations) is due to strong initial weight updates performed by GDP.

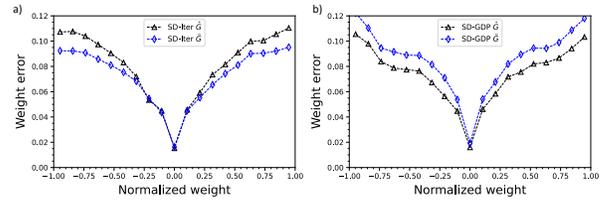

Fig. 6. We compute the weight error (y-axis) between the target conductances ($G$) and the readout conductances ($\tilde{G}$), as well as the estimated conductances ($\hat{G}$) and compare iterative programming (a) to GDP (b). Iterative programming leads to a lower distance between $\tilde{G}$ and $G$ than GDP, but higher distance between $\hat{G}$ and $G$. Hence, GDP finds a set of readout weights that actually produces a larger error w.r.t. the target weights, but still yields lower MVM error. Purposefully programming the weights "away" from the target weights indicates that GDP accounts for small nonlinearities in the tile.

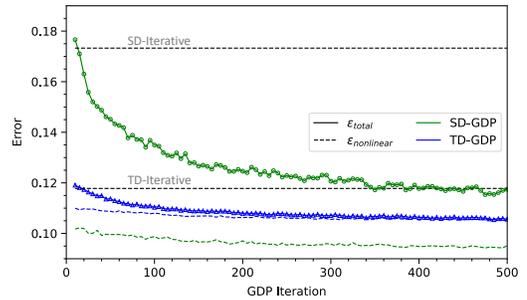

Fig. 8. Illustration of $\epsilon_{total}$ and $\epsilon_{nonlinear}$ for single-device (SD) and two-device (TD) GDP for 500 iterations. When GDP is used with two devices, the gap between the total and nonlinear error is closed. GDP with only one device reaches the two-device iterative baseline.

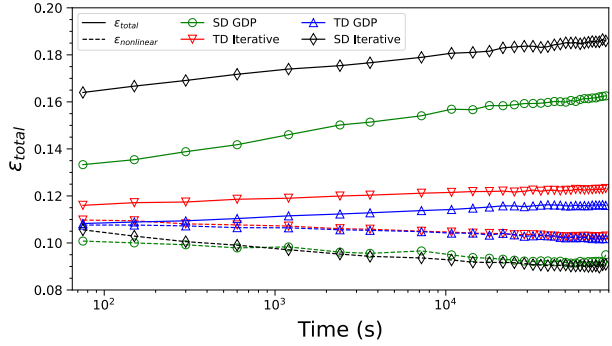

Fig. 9. The effect of drift variability on the total (solid) and non-linear (dashed) error over time (24 hours). Note that the non-linear error is higher for the methods that use two devices (red & blue) because the current on the ADCs is doubled and therefore produces more nonlinearities.

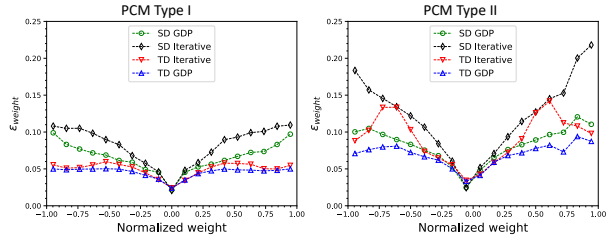

Fig. 11. Weight error for two types of PCM devices, where PCM II has lower conductance than PCM I. SD and TD GDP outperform iterative programming for both types of PCM devices, especially for PCM II where iterative programming leads to large errors due to inaccurate reads of the low conductance values that cannot be resolved by the ADC.

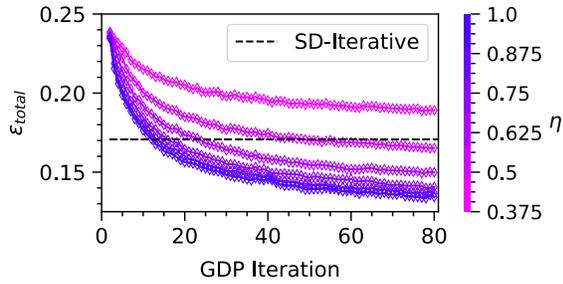

Fig. 13. Illustration of the MVM error over the optimization steps for GDP. The black dashed line shows the best performance of the iterative approach. We generally found that the exact value of the learning rate $\eta$ (color coded) is not of high importance, but rather that one should avoid too small values.

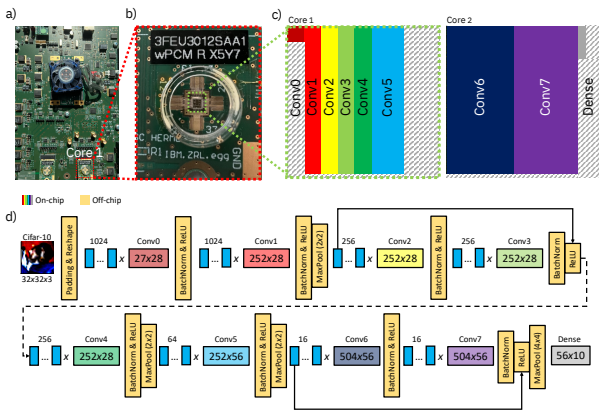

Fig. 15. The experimental platform (a) [7] comprises two PCM cores (b) with $256 \times 256$ unit cells. The weights of the scaled down ResNet-9 architecture (d) are mapped to both cores as shown in (c).

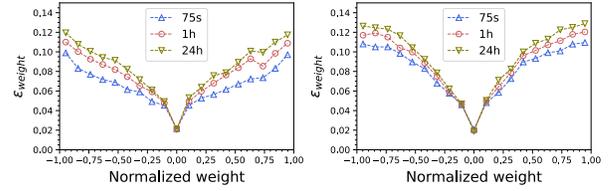

Fig. 10. Comparison of the weight error between SD-GDP (left) and SD-Iterative (right) over a time-span of 24h.

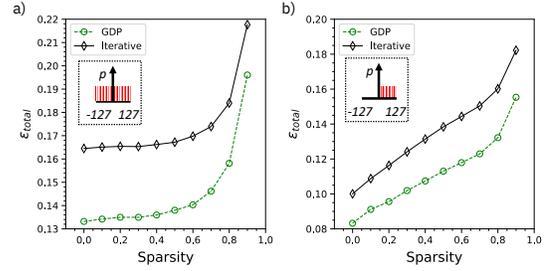

Fig. 12. (a) Improvement of GDP (green) over the iterative method (black) persists when performing MVMs using inputs of varying levels of sparsity. (b) This also holds when the input distribution (inset) is changed. The errors tend to get larger with increasing sparsity due to lower SNR.

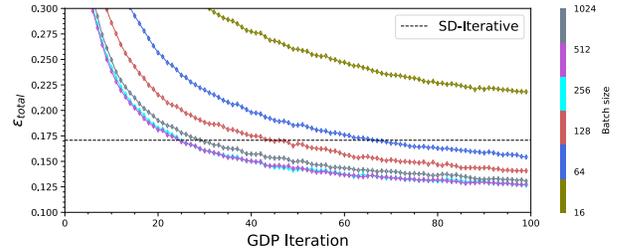

Fig. 14. GDP begins to outperform the baseline (black dashed line) at a batch size of 64 and performs best for batch sizes greater or equal to 256.

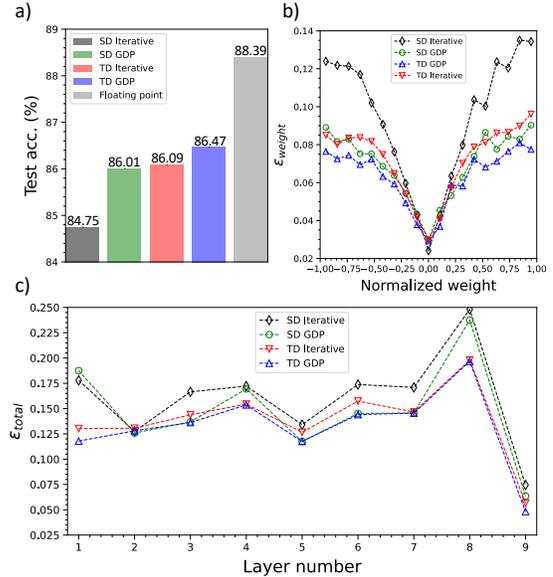

Fig. 16. Decreasing the MVM error has a direct effect on downstream tasks such as neural network inference. (a) Single-device (SD) and two-device (TD) GDP significantly boost inference accuracy on a scaled down version of ResNet-9. This increase in performance is a result of reducing the weight error (b), which is also reflected in the per-layer total error (c). For each method, we programmed the network weights and ran inference on the experimental platform one time on the full 10k images of the CIFAR-10 test dataset.